\documentstyle[prb,aps]{revtex}

\begin{document}

\title{Role of  phason-defects on the conductance of a 1-d  quasicrystal}
\author{ K. Moulopoulos and S. Roche}
\address{ Laboratoire d' Etudes des Propri\'et\'es Electronique
des Solides, CNRS,  38042 Grenoble\\
e-mail : roche@lepes.polycnrs-gre.fr}
\date{ June 95}

\maketitle

\begin{abstract}

We have studied the influence of a particular kind of phason-defect on the
Landauer resistance of a Fibonacci chain.
Depending on parameters, we sometimes find the resistance to decrease
upon introduction of defect or temperature, a behavior that also appears in
real
 quasicrystalline materials.
We demonstrate essential differences between a standard
tight-binding model and a full continuous model.
In the continuous case, we
study the conductance in relation to the underlying
chaotic map and its invariant. Close to conducting points, where the invariant
vanishes,
and in the  majority of cases studied, the resistance is found
to decrease upon introduction of a defect.
Subtle interference effects  between a sudden phason-change
in the structure and the phase of the wavefunction are also found,
and these
 give rise to resistive behaviors that produce exceedingly simple
and
regular patterns.
\end{abstract}

\section{ Introduction}

According to recent experiments\cite{aussois}, quasicrystals have curious
(for metallic materials)
 transport
properties.
For example,
anomalously high values of the low-temperature resistivity have been reported,
and the resistivity goes down with introduction of defects
or with increase of temperature.
The anomalously high resistivity has
 been partly accounted for by the existence of a pseudogap at  the Fermi
energy.
For the temperature- or defect-dependence there is no clear consensus yet.
 In numerical simulations of 2-d systems,
fluctuating behavior for the resistance of a defected
(with $\it{random}$ disorder) with respect to that of a  pure system is
observed,
namely the resistivity can either go up or down\cite{fujiwara}.

Here we attempt to address these behaviors
for a 1-d system
by focusing on the electronic $\it{phase\;relations}$ in real space
 that actually determine
the conductance.
We  study, for example,   the scattering of electrons on a quasiperiodic
(Fibonacci)
arrangement of $\delta$-function potentials (a continuous model
with the full phase coherence included).
 After determining the
Landauer conductance of a finite part of such a system, then we
introduce
 a particular type of defect of a step-form in hyperspace\cite{janot}
(called ``phason-defect" from now on) and we study how  this defect
$\it{influences\;\; those\;\; phase\;\; relations}$.
This is therefore a study of the effect of an abrupt ``phase"-change
in the structure (with ``phase" defined in hyperspace)
on the phase-coherence of the electronic wavefunction
in real space (with phase defined
in the usual quantum-mechanical way).
(The role of the initial ``phase" of pure (undefected) chains on the
resistance has been studied by other authors\cite{kubo}).

The quasicrystalline structure can induce exotic (critical)
 states\cite{kohmotospec}
that are intermediate between localized and extended states.
Especially in the pure Fibonacci chain all states are critical
independent of values of parameters.
One consequence that we observe,
 for a $\it{finite}$  part of the Fibonacci chain, is
a rather unpredictably fluctuating  variation of the Landauer resistance
with the length of the sample\cite{kohmotoresi}
(except when we are on special energy-regions
of integrability, as we will see below).
But in addition to this  irregular
variation for a pure Fibonacci chain,
we here also attempt the first study of the variation of the Landauer
resistance
for the phason-defected system as well, both with respect to the length of the
sample
and
also with respect to the position of the defect.
The comparison of the resistances between the pure and the defected
system shows some interesting  regular patterns, provided that we are close to
special points associated with the vanishing of the invariant $I$
of the underlying dynamical map\cite{kohmotospec}
(on which points, both problems are integrable).
In addition, in the majority of values of the parameters,
the pure system is found to be
more resistive than the defected one.

It is also demonstrated that our use of a continuous model is
critical in obtaining our results. As we show, these results would have
actually been
missed had one first made
 a tight-binding approximation (i.e. keeping only nearest-neighbor
overlaps).
 Our work is therefore a concrete
example of the danger\cite{baake} that  the usual discretized  approximations
incorporate,
especially when systems with  quasiperiodicity are considered,
where subtle interference effects between the phase of the
wavefunctions and the ``phase" in hyperspace can be expected.

\section{ Tight-Binding Model}

Let us first use
 a tight-binding  model, in order to compare with the fuller treatment
that is given later.
For simplicity we fix the Fermi (or incidence) energy in the center of the band
of the outside conductor (coupled to the right and left of our
finite system).
We then study the scattering of an electron coming from the left with
a wavevector $k$, and we determine the Landauer resistance\cite{landauer}
of our finite system
 (ratio of the reflection to the transmission coefficient).
The method used for this is the standard transfer matrix
method\cite{kohmotoresi}.

\vspace{5pt}

The Hamiltonian for a general (mixed) tight-binding model is
$$H=\sum_{n}\epsilon_{n}\mid n><n\mid +\;\;t_{n}\Bigl(\mid n><n+1\mid +\mid
n+1><n\mid \Bigr) \label{ham}$$
with ${t_N}={t_0}=t_{out}$ and $E-\epsilon_{out}=2t_{out}\cos ka$, where
  $k$ is the  incident wavevector and $a$ is the lattice constant of
the outside conductor
(``out"). \smallskip
The Schr\"odinger equation then reads in matrix language

$$\pmatrix{
\Psi_{n+2} \\
\Psi_{n+1} \\
}=M_{n}.\pmatrix{
\Psi_{n+1}\\
\Psi_{n}\\}$$

$$M_{n}=
\pmatrix{
{E-\epsilon_{n+1}}\over{ t_{n+1}}&-{
t_{n}\over{\strut  t_{n+1}}}\\
{1}&{0}\\
}. $$

The Landauer resistance (for spin-${1 \over 2}$ fermions)
is then, in atomic unis ($\pi \hbar/e^2$)

\begin{equation}
{{ \cal R}\over{\cal T}}={\mid
T^{12}_{N}\mid}^{2}\\
\end{equation}

with the matrix $T_{N}=\Theta^{-1}.S^{-1}.P_{N}.S$, where $P$ is the
total transfer matrix, namely
$P_{N}=M_{N-1}.M_{N-2}...M_{0},$
and $\Theta$ and $S$ are defined by

$$\Theta=\pmatrix{
e^{-ikNa}&0\\
0&e^{ikNa}\\
}$$
$$
S=\pmatrix{
e^{-ika}&e^{ika}\\
1&1\\
}$$

\vspace{5pt}

\subsection{ Example : Periodic system}
\vspace{5pt}

In this case
$t_{1}=...=t_{N-1}=t_{N}\equiv t$ and
$\epsilon_{1}=\epsilon_{2}=...=\epsilon_{N}\equiv \epsilon$. Then
$P_{N}=M^{N}$,
and two cases appear naturally :

\vspace{5pt}

\begin{eqnarray}
a)\ \mid {{E-\epsilon}\over{2t}}\mid <1\Longrightarrow
{{ \cal R}\over{\cal T}}
&=&{{\biggl({{ \epsilon_{out}-\epsilon}\over
2t}\biggr)^2}\over{\sin^{2}ka}}\\
& &{{\sin^{2}N\phi}\over{\sin^{2}\phi}} \hbox{with $\phi$ defined by}\\
 \cos\phi&=&{{E-\epsilon}\over2t}\\
\end{eqnarray}

\begin{eqnarray}
b) \mid{{E-\epsilon}\over{2t}}\mid>1\Longrightarrow
{{ \cal R}\over{\cal T}}
&=&{{\biggl({{\epsilon_{out}-\epsilon}\over
2t}\biggr)^2}\over{\sin^{2}ka}}\;{{\sinh^{2}N\phi}\over{\sinh^{2}\phi}}\\
\hbox{with $\phi$ defined by}\ \cosh\phi&=&{{E-\epsilon}\over {2t}}\\
\end{eqnarray}

\vspace{5pt}

\smallskip

Case (a) gives  oscillatory behavior of the resistance  with size, while
case (b) gives  exponential increase for large size. In the intermediate
 case
${{E-\epsilon}\over{
2t}}\rightarrow 1$ we have $\;\; {{ \cal
R}\over{{\cal T}}} ={{ ({{
\epsilon_{out}-\epsilon}\over { 2t}})^{2} }\over {
\sin^{2}ka} }\  N^2,$ i.e. the resistance grows as the square of length.
In the limit $N \rightarrow \infty$ case (a) corresponds to allowed energies
(bands),
while case (b) to forbidden ones (gaps).

One can also trivially solve a periodic system with more complex unit cell (e.g
quasicrystalline approximants). In this case the phase $\phi$ depends on
the  internal
structure of the unit cell and is a more complicated function of various
parameters, but straightforward to determine for a given unit cell.

Finally, the introduction of even a single defect leads to interference effects
that can easily  be
studied through this method. It is found\cite{costas} that the behaviour
in the presence of a  defect seems to be correlated with
whether we are in case (a) or (b) above.

\smallskip

\subsection{  Fibonacci and a Phason-defected system}

\vspace{5pt}

Take for simplicity an off-diagonal model ($\epsilon_{n}=\epsilon_{out}=0$).
Also set $E=0$.
 Hopping elements $t's$ are forced to take two values $t_{a}, t_{b}$ arranged
in :

\vspace{5pt}

1) a Fibonacci way
\smallskip
2) with a phason-defect (corresponding to a ``phase"-change of $2/\tau$
in hyperspace , with $\ \tau
={{\sqrt{5} +1}\over2}$ the golden mean), but in both cases in such a way as to
comply with
boundary conditions $t_{N}=t_{0}=t_{out}=t_{b}$. This imposes restrictions on
the possible values of N :
\smallskip
N=5,13,18,26,34,39,47,52,60,... (Their difference is seen to be a Fibonacci
arrangement of the numbers $8$ and $5$).  For example, for  $N=39$ we have
\begin{eqnarray}
\hbox{Pure}&:&(b)abaababaabaababaababaabaababaabaababaa(b)\\
2/\tau-\hbox{shifted}&:&(b)aababaabaababaababaabaababaabaababaaba(b)\\
\hbox{Defected}&:&(b)abaababaabaaba{\bf bb}abaabaababaabaababaaba(b)\\
\end{eqnarray}
where in the last line  we show only one of the possible places where the
defect (i.e. an abrupt ``phase"-change of ${2/\tau}$ in the structure;
see Fig.1) can be
introduced.
\vspace{5pt}

We now give an analytic solution for the
Landauer resistances of the two systems (pure and defected) :
\medskip
\vspace{5pt}

For a given $N$ we find that
\begin{eqnarray}
\Bigl({{ \cal R}\over{\cal
T}}\Bigr)_{pure} &=&\Bigl({  { \gamma^{s}-\gamma^{-s}}\over
{ 2}}\Bigr)^{2}\\
\Bigl({{ \cal R}\over{\cal
T}}\Bigr)_{defected} &=&\Bigl({  {\gamma^{s-1}-\gamma^{-s+1}}\over
{ 2}}\Bigr)^{2}\\
\end{eqnarray}

\vspace{5pt}

with ${ \gamma ={t_{a}\over t_{b}}}$
and $s$  an integer power (of both signs)
 that is an irregularly fluctuating function of the length  $N$
(around the value $s=0$ corresponding to complete transparency).
\smallskip
 We conclude  (for $\gamma >1$ ) that for $s>0
\rightarrow \Bigl({{ \cal R}\over{\cal
T}}\Bigr)_{defected}<\Bigl({{ \cal
R}\over{\cal T}}\Bigr)_{pure}$.
We see that the appearance of  such cases (which we will call ``favorable")
depends on the behavior of the pure chain with  length (this behavior
fluctuates
in an irregular fashion).

 For big values of $N$   we get 50\% of
this favorable  behaviour. (We have also carried out cases of multi-defected
sequences\cite{roche}, where we also find irregularly fluctuating behavior,
but
with the {\it scale} of fluctuations being much higher).

\section{
 Continuous Schr\"odinger equation models}

\vspace{5pt}

In this continuous case, which has the advantage of not suffering from
truncation approximations,  the matrix form of the Schr\"odinger equation
results
from matching wavefunctions and derivatives at scattering points.
This yields a generalized
 Poincare map\cite{spanish2} of the following form

$$\pmatrix{\Psi_{n+1}\\
\Psi_n\\}
= M_n .
\pmatrix{\Psi_{n}\\
\Psi_{n-1}\\}$$

with
$$M_n=\pmatrix{K_{11}(n+1)+K_{22}(n)
{
{{K_{12}(n+1)}\over{K_{12}(n)}}
}
&
{
-{{K_{12}(n+1)}\over{K_{12}(n)}}
}
\\
{1}&{0} }$$
\vspace{5pt}

In the special case of a scattering potential of the form
$V(x)={ \sum_n \lambda_n \delta(x-x_n)}$ and with
$k=\sqrt{{{2 m E}\over {\hbar^2}}}$ one obtains\cite{Poincar1}
$$K_{11}(n+1)=\cos k (x_{n+1}-x_n)+
{{\lambda_n}\over {2 k}} {{2 m}\over {\hbar^2}}\; \sin k (x_{n+1}-x_n)$$
$$K_{12}(n+1)
=
{{\sin k (x_{n+1}-x_n)}\over k}$$
$$K_{22}(n+1)=
\cos k (x_{n+1}-x_n)+
{{\lambda_{n+1}}\over {2 k}} {{2 m}\over {\hbar^2}}\; \sin k (x_{n+1}-x_n).
$$
\vspace{5pt}

\subsection{Comparison with Section 2}
\vspace{5pt}

Let us pause for a moment to see
how a tight-binding approximation is usually derived:
To change to a localized description one typically  writes the wavefunction
$\Psi(x)$ in terms of localized states $\phi_n$, namely
$\Psi(x)={ \sum_n C_n \phi_n(x-x_n)}$
with
$\phi_n(x-x_n)= \sqrt{\lambda_n} e^{-\lambda_n |x-x_n|}$
and then neglects the overlap between distant $\phi_n$'s;
assuming that only nearest-neighbor overlaps are significant
one  obtains a tight-binding  model with site and hopping
elements\cite{spanish}
$$\epsilon_n=-{1 \over 2} \lambda_n^2$$
$$t_{n,n \pm 1}=-\sqrt{\lambda_n^3 \lambda_{n \pm 1}} e^{-\lambda_{n \pm 1}
|x_{n \pm 1}-x_n|}.$$

It is important to note then that in the special case of
  a continuous model with identical
strengths $\lambda$ and quasiperiodic arrangements ($\Delta x$'s taking  two
values
in a Fibonacci sequence), which  we later analyze
in detail (see Subsection B below), we would just obtain a simple off-diagonal
tight-binding approximation
(since in this case all the $\epsilon_n$ are identical). This demonstrates a
defficiency of the
tight-binding approximation, on  which we now elaborate:

\vspace{5pt}

In the case of Fibonacci arrangements (let us say of both $\epsilon$'s and
$t$'s) and for $N=F_n$ (a Fibonacci number) we have
in the tight-binding approximation (with definitions
$P_n \equiv M_{F_n}...M_1$, with $M$'s given in
the well known\cite{kohmotospec}
recursive scheme $P_{n+1}=P_{n-1}.P_n$, with starting matrices
$\;\;\;P_1=\pmatrix{ {{E-\epsilon_a}\over {t_a}} & -1\\ 1 & 0 \\}\;\;$, and
$\;\;\;P_2=\pmatrix{ {{E-\epsilon_a}\over {t_a}} & -{{t_b}\over {t_a}}\\ 1 & 0
\\}.
\pmatrix{ {{E-\epsilon_b}\over {t_b}} & -{{t_a}\over {t_b}}\\ 1 & 0 \\}\;\; $
and hence the usual trace map\cite{kohmotospec} $x_{n+1}=2 x_n x_{n-1} -
x_{n-2}$
(with the definition  $x_n={1 \over 2} tr(P_n)$).
This map has the well-known  invariant $I=
x_{n+1}^2+
x_{n}^2+
x_{n-1}^2-
2 x_{n+1}
x_{n}
x_{n-1} -1 $.
Straightforward evaluation  for the above case yields
$$I= {1 \over 4} \Biggl(
(\epsilon_a-\epsilon_b)
\biggl(
{{E-\epsilon_b}\over {t_b^2}}-
{{E-\epsilon_a}\over {t_a^2}} \biggr)
+
\biggl( {{t_a}\over {t_b}}
 -{{t_b}\over {t_a}}
\biggr)^2
\Biggr)$$

\vspace{5pt}

In our case of a simple off-diagonal model (i.e. $\epsilon_a=\epsilon_b $)
(but also for the case of  diagonal models  (i.e. $t_a=t_b $))
I is  E-independent as seen from , always positive,
and it never vanishes. This is a deficiency of the approximation, as discussed
below. (Note that even in the case of a mixed model, where $I$ vanishes only
for a single value of $E$, the elementary matrices $M$ do $\it{not}$
commute at this value, which is a major difference with the full continuous
model,  as will be discussed in Subsection B).
\vspace{5pt}

We will see below that,
because of implicit truncation errors, the tight-binding models
miss interesting patterns associated with the $\it{zeros}$ of the invariant $I$
of the underlying dynamical map, where the basic matrices commute.
These are  actually relevant to conduction and are treated next
(in Subsection B) in the continuous
Schr\"odinger formulation.

\vspace{5pt}

\subsection{ Example  : Periodic System}

Let us first discuss a problem where the scattering potential is
an array of
equally-spaced (with length $a$) $\delta$-functions of
equal strengths ($\lambda$).
Then, again two cases appear naturally (we take ${2m\over {\hbar^2}}=1$,
or, equivalently, replace everywhere $k$ by $k* {\hbar^2 \over {2 m}}$
in what follows):
\medskip
\medskip
\vspace{5pt}

a) For $\mid \cos ka+{\lambda\over 2k}\sin ka\mid<1\Longrightarrow
\Bigl({{ \cal R}\over{\cal
T}}\Bigr)={({\lambda\over 2k})}^2\times
{ {\sin^{2}N\phi}\over{ \sin^{2}\phi}} $
\smallskip
\noindent
with $\phi$ defined by  $\cos\phi=\cos ka+{\lambda\over 2k}\sin ka$ (which
is seen to be the same as in the  usual treatment of the
Kr\"onig-Penney model\cite{KP1} for the case of allowed bands (for
$N\rightarrow\infty$)).
We note again the oscillatory behavior of the resistance with length.

\vspace{5pt}

It is easy to show that the naturally appearing phase $\phi$  determines the
crystal momentum $q$
through $\phi=q a$ (with such an identification, the resulting wavefunctions
indeed satisfy the Bloch theorem, as can be easily checked).
We can  also  see that the
 zeros of $\Bigl({{ \cal
R}\over{\cal T}}\Bigr)$
determine the allowed values of
$q$ (from band theory) for the $\it{infinite}$
periodic system. This is an important point, because it motivates
our later treatment of a Fibonacci problem, where we will enforce the
vanishing of the resistance in order to simulate the thermodynamic limit.
\medskip
\medskip

b) For  $\mid \cos ka+{\lambda\over 2k}\sin ka\mid>1\Longrightarrow
\Bigl({{ \cal R}\over{\cal
T}}\Bigr)={({\lambda\over 2k})}^2\times
{ {\sinh^{2}N\phi}\over{ \sinh^{2}\phi}}
$
\smallskip
\noindent
with $\phi$ defined by  $\cosh\phi=\cos ka+{\lambda\over 2k}\sin ka$. This
corresponds to the case
of gaps in the limit $N\rightarrow\infty$. Indeed the resistance grows
exponentially
as $N\rightarrow\infty$. In this case, wavefunctions
 decrease with a characteristic length $\Lambda={
2aN\over{ \ln (1+{{ \cal
R}\over{\cal T}})}}$
\medskip
Once again the study of complex unit cells is of course possible.
Also the introduction of a single defect results in
 interference effects that can be easily studied through this method.

\vspace{5pt}

\subsection{ Schr\"odinger equ. with
Fibonacci and  Phason-defected $\delta$-potentials}

\vspace{5pt}

In the pure case  a recursive procedure of the form
$P_{n+1}=P_{n-1}.P_n$ can also be established
for this case  (with starting matrices depending on
the description).
For example, in the case of equally spaced ($a$) $\delta$-potentials
with $\lambda_n= \{\lambda_a,\lambda_b \}$ in Fibonacci arrangement,
a natural description is the one given in the  previous section
(in terms of matrices $M_n$)
(with $P_n= M_{F_n}...M_1$). Then the invariant  is
$I={{(\lambda_a-\lambda_b)^2 \sin^2(k a)}\over {4 k^2}}$
i.e. $I \geq 0$. (Its zeros constitute a periodic pattern $k_s={{n \pi}\over
a}$, ($n \neq 0$)).
\vspace{5pt}

$\underline{Our \;\;system}$: The system, however,  we will analyze in detail
corresponds
to $\delta$-potentials with
 equal strengths ($\lambda$) but quasiperiodic arrangements
$\{(x_n-x_{n-1}) \}=
\{a,b \}$
in Fibonacci arrangement\cite{suto,KP2,continu_spect}, as already mentioned
earlier.
\vspace{5pt}

 In this case, an alternative description
(in terms of the  coefficients $A_n$ and $B_n$ of the two linearly
independent  plane waves in the region between two $\delta$ potentials)
is more natural. One obtains\cite{suto}
$$\pmatrix{A_{n+1} \\ B_{n+1} \\}=
\Lambda(n).
\pmatrix{A_{n} \\ B_{n} \\} \label{contmatrixSchr} $$
with
$$\Lambda(n)=
{
\pmatrix{
(1-{{i \lambda}\over{2 k}} ) e^{i k (x_{n+1}-x_n)} &
-{{i \lambda}\over{2 k}}  e^{i k (x_{n+1}-x_n)} \\
{{i \lambda}\over{2 k}} e^{-i k (x_{n+1}-x_n)} &
(1+{{i \lambda}\over{2 k}} ) e^{-i k (x_{n+1}-x_n)}\\}
}
$$
 for $\delta$-potentials (once again we have set $2 m/\hbar^2=1$).
In this description the Landauer resistance is  given again by
${\cal R \over \cal T}= |P_{12}|^2$ with $P$ the product of $\Lambda$'s, namely
 $P_n= \Lambda(F_n) ...\Lambda(1)$. In this case the invariant $I$ of the
underlying dynamical map   turns out
to be\cite{invariant}
$I={{\lambda^2 \sin^2k(a-b)}\over {4 k^2}}$ i.e. also $I \geq 0$.
Its zeros now constitute a periodic pattern $k_s={{n \pi}\over {a-b}}$
and will be the focus of our work in what follows.

\vspace{5pt}

These special points $k_s$ are missed in the corresponding
 tight-binding model, as we showed
earlier.
In the present  continuous model they turn out to  correspond to commuting
consecutive matrices, i.e.
$[P_{n+1},P_n ]=0\;\;$ (which is consistent with the known
relation\cite{baake}
 $4 I+2=tr (P_n.P_{n+1}.P_n^{-1}.P_{n+1}^{-1})$), but even stronger,
they lead to commuting elementary matrices \ref{contnewmatrix}. Indeed,
it turns out that
$$\left [ \Lambda (\Delta x=a),
 \Lambda (\Delta x=b) \right ] =
\lambda \; {{e^{i k (a-b)}}\over {4 k^2}}\; (1-e^{2 i k (b-a)})
\pmatrix{
\lambda & \lambda-2 i k \\
-\lambda-2 i k& -\lambda\\
} $$
which vanishes for all special points $k=\{k_s \}$.
\vspace{5pt}

 One can therefore say that,
from a conduction point of view,  the problem looks
formally similar to a periodic problem. This is more concretely described
through the following properties of the special points $\{k_s \}$:

\vspace{5pt}

1) They are the conducting points (extended states) surviving\cite{suto}
 in the limit
$N \rightarrow \infty$.
\vspace{5pt}

2) The Landauer resistance can be written $\it{exactly}$ in closed form
for $k= \{k_s \}$, and the expression looks like that
of a periodic system (see below).

3) The set of points $k_s$ is robust\cite{baake} against disorder
(essentialy because of the above mentioned commutations).
In fact it exists either for a periodic or  for a random system (of two
letters).

\vspace{5pt}

In what follows, we focus on these points
$\;\;\;k_s={{n \pi}\over{a-b}}$ with
$a=\tau={{1+\sqrt{5}}\over 2}$,$\;\;b=1$ (Fibonacci chain):
Exactly on those points we get the Landauer resistance in closed form
(for the Fibonacci or any disordered system of $\{a,b\}$):
$${\cal R \over \cal T}|_{_{_{k=k_s}}}=\biggl({\lambda \over {2 k_s}} \biggr)^2
\;\;{{\sin^2 N \phi}\over{\sin^2 \phi}} \label{resspecial}$$
with $\phi$ defined by  $|\cos \phi|=|\cos k_s a+
{\lambda \over {2 k_s}} \sin k_s a |=
|\cos k_s b+
{\lambda \over {2 k_s}} \sin k_s b | $ (for $2m/\hbar^2=1$).
\vspace{5pt}

Motivated by our earlier discussion on a finite piece of a
 periodic system (where we saw that the
values of $k$ that made the oscillating resistance $\it{vanish}$ correspond to
the
allowed states of the  $\it{infinite}$ system),
we now $\it{enforce}$ a conducting behavior (resonance):
For any fixed $N$, we choose $\lambda$ in such a way
as to have a vanishing resistance (exactly at $k=k_s$), namely
${\cal R \over \cal T}|_{_{_{k=k_s}}}= 0 $
(for $\it{both}$ the pure Fibonacci and the phason-defected system).
(This  gives  the representative behavior in the thermodynamic limit
where this vanishing is actually expected for $k = k_s$ and for any $\lambda$;
see below).
As a consequence we get $(N-1)$ different appropriate values of $\lambda$
(both positive and negative) given by
$$\lambda_s= {{2 k_s (\cos \phi_s-\cos k_s b )}\over {\sin k_s b}}
\label{lambdas}$$
with the internal phase $\phi_s$ defined by
$\phi_s={{m \pi}\over N}$
with $m=1,...,N-1$ that correspond to vanishing
${\cal R \over \cal T}|_{_{_{k=k_s}}}. $
The $(N-1)$ values of the phase $\phi_s$ are symmetrically placed
around ${ \pi \over 2}$ and as $N$ increases they cover densely
the entire upper half of the trigonometric circle. Consequently,  the
vanishing of
${\cal R \over \cal T}|_{_{_{k=k_s}}} $
{\it is} representative of the behavior in the thermodynamic limit
for $\it{any}$ arbitrary  value of $\lambda$,
and that is the reason we enforce it even for a finite system.
By doing so, we then study
 the behavior
  $\it{around}$ the  conduction points
 $\;k_s={{n \pi}\over {\tau-1}}$. Below we summarize our results
on the behavior of the resistance, of both the pure and the defected system,
as $k$ varies in the local neighborhood of the conduction points $k_s$:
\vspace{5pt}

1). Smallest system (consistent with boundary conditions): $N=5$
\vspace{5pt}

The results are given in Fig.2, where we show
the local regions around the lowest conducting point
($n=1$) for one value of $\phi_s$, showing that the resistance of the
defected system $\it{always}$ $\it{rises}$ $\it{lower}$
 than the one of the pure system (this is also true for {\it any} $\phi_s$).
We also show the complicated behavior of the Landauer resistance in
more extended regions around a special point and for one particular $\phi_s$.
\vspace{5pt}

Of course the problem is analytically solvable. By way of an example we give
the
 analytical solution
for $n=1$ ($k_s={{\pi}\over {\tau-1}}$), $\;\;m=1$ ($\phi_s={\pi \over 5}$)
which is
$${\cal R \over \cal T}|_{_{_{pure}}}= 45.3087\;\;\Biggl(k- {\pi \over
{\tau-1}}
\Biggr)^2 +...$$
$${\cal R \over \cal T}|_{_{_{defected}}}= 30.7868\;\;\Biggl(k- {\pi \over
{\tau-1}}
\Biggr)^2 +...$$
\vspace{5pt}

This is the typical behavior for $\it{all}$ points examined
($1 \leq n \leq 5$, all $m$'s)
i.e. the Landauer resistance of the phason-defected system
around the conducting points $k_s$, rises (from zero)
$\it{always}$ $\it{lower}$ than the corresponding one for the pure system.

2).$N=13$

Extensive numerical results have been obtained in this case,
where 	 we have $3$ possible defected systems,
 for $1 \leq n \leq 5$ and for all $12$ phases $\phi_s$.
Out of those $12$ cases, in $8$ of them $\it{all}$ three defected systems
have Landauer resistances $\it{lower}$ than that of the pure system.
 In
only  $2$ cases
one defected sequence rises slightly higher
and also in $2$ other  cases another sequence  also rises slightly higher
than the pure system.
We call these few cases (where the defected system is more resistive)
 ``unfavorable cases".

\vspace{5pt}

3) Long chains
\vspace{5pt}

We have carried out numerical calculations on chains up to more
than 3000 sites, taking as a numerical  convergence criterium the unitarity
condition of our total transfer matrices.
\vspace{5pt}

Because of the full phase coherence, the transmission
behavior of the defected chains depends on the discrete position $x_P$ of
the defect in interesting ways. In long chains the variable $x_P$
becomes quasi-continuous, and the discrete and asymmetric patterns found
for small chains (see for example Fig.3 and Fig. 4-c), now appear smooth and
symmetric
(Fig.4-a and Fig.4-b). Moreover, their actual
form depends on the value of  $\phi_s$, which can be considered as a
discrete  label parametrizing the
family of potential strengths $\lambda$ that enforce vanishing of resistance on
specials
points. The results show interesting patterns summarized below and
shown pictorially in Fig.5.
\vspace{5pt}

 We observe  a symmetry in the qualitative behavior
 around the value  $\phi_s ={\pi \over 2}$ (although the values of $\lambda$
corresponding
to symmetric values of $\phi_s$ are always different, as seen from equ.
\ref{lambdas}.
Up to this  symmetry, we also observe a type of recurrent simplicity
in the resistive behavior of the defected system, which we loosely call
``cyclic" (with period $P$).
For small to medium values of $\phi_s$ (in more detail,
for $m  <  N-4P$, with $m$ defined by
$\phi_s = m {\pi \over N}$ and with $P$ being  the number
of possible defect-points (note that for a given chain-length $N$ there is a
unique $P$
always satisfying the inequality $N  <  5 P$ for any $N >5$), we observe
{\it exceedingly} simple oscillatory patterns (Fig.4).
Furthermore,  we observe that in these cases
the defected
system is {\it always} less resistive than the pure one.
\vspace{5pt}

A qualitatively similar behavior is  observed in the mirror
symmetric region (see the two shaded regions in Fig.5).
It is interesting to point out that the above ``favorable" behavior
is valid for sufficiently strong absolute values of $\lambda$, namely
$| \lambda | \geq \lambda_0$ with $\lambda_0={{2 \pi}\over {\tau-1}}\;
{{|cos{( {{4 P  \pi}\over N})}| \pm cos{({\pi \over {\tau-1}})}}
\over {|sin{( {\pi \over {\tau-1}})}|}}$,
with the lower sign corresponding to the 1st  and the upper sign
to the 2nd quadrant of the upper half of the trigonometric circle.
As $m$ increases towards $P$ we observe additional fluctuations
resulting in a rather regular modulated pattern,
containing a finite and small number of harmonic modes (Fig.6)
The above mentioned ``cycles" in the recurrence of the simplest
behavior, occur whenever
$m$ approaches $n P$ (again up to the symmetry around $\phi_s = \pi/2$).
As we cross  the cycles we also observe abrupt changes of ordering in the
resistance of the defected with respect to that of the pure chain (Fig.7).
Finally, close to the symmetry point $\phi_s = {\pi \over 2}$ we  observe
largely fluctuating and modulated patterns that look self-similar
(Fig.8). These resistive patterns become chaotic (Fig.8(c))
when we move sufficiently far from the special integrability points.
\vspace{5pt}

(The above numerical observations are stable upon approaching the special
point (i.e. decreasing $\epsilon$) up to the lowest value of
$\epsilon$ where our convergence criterion is satisfied.
\vspace{5pt}

We view the above results (summarized pictorially in Fig.5)
 as interference effects between the abrupt
``phase"-change in hyperspace (that occurs at some point in physical space),
and the phase of the wavefunction at that point. The
Landauer resistance viewed as a  function of the discrete variable
 $x_{P}$ (that actually corresponds to a family of different physical
systems) is here seen to carry a ``memory" of the hyperspace
by showing a type of ``hyper-coherence", that depends on physical
parameters in rather subtle ways.

\vspace{5pt}

\section{
Comment on introduction of finite Temperature $T$}.

\vspace{5pt}

The Landauer resistance at finite temperatures is given by\cite{anderson}
$${\cal R \over \cal T} (T,\mu)=
{{
\int (-{{\partial f}\over {\partial E}} )
\biggl( 1- T(E) \biggr) dE
} \over
{\int (-{{\partial f}\over {\partial E}} )
 T(E)  dE}}$$
with
$f(E)={1 \over {1+e^{\beta (E-\mu)}}}$, $\;\;\;\beta={1 \over {k_B T}}$.

\vspace{5pt}

For $N \gg 1$ the transmission coefficients $T(E)$'s are dominant\cite{suto}
in the local regions around the special points $k_s$
(studied on last pages).
Because of the smearing of the Fermi function at low temperatures,
the behavior of the Landauer resistance with temperature
depends on the location of the Fermi energy $\mu$.
If $\mu$ is close to a special energy
(i.e. if $\mu \sim {{\hbar^2 k_s^2}\over {2 m}}$)  then
the decrease or increase of ${\cal R \over \cal T}$ of the defected system
compared
to the pure system, depends on whether this particular $k_s$ is a favorable or
unfavorable
point in the sense discussed on the previous pages.
We conclude therefore, from our results above,
 that, statistically speaking, in the  majority of cases
the resistance will decrease upon increase of temperature.

\vspace{5pt}

\section{ Conclusions}
\vspace{5pt}

Motivated by the anomalous transport properties of quasicrystals
with defect and temperature, we
 have analysed the role of phason-defects on the Landauer resistance
of a finite Fibonacci chain.
In a tight-binding model  with incident energy being fixed
 at the center of the band,
the resistance is modified in an irregular fashion
and in accordance with  the behavior of the pure
Fibonacci chain with the size of the system, with both positive and negative
effect appearing statistically  in equal percentage.
In a fuller treatment of the continuous Schr\"odinger equation with
$\delta$-function potentials, the modification of the resistance
depends on the location of the Fermi level with respect to special
energies corresponding to extended states, and also on the position
of the defect.
For finite chains, the majority of cases studied show a decrease of the
Landauer resistance upon introduction of defect and temperature.
\vspace{5pt}

By comparing the two models, we have noted that continuous models,
being free from truncation errors,
preserve the phase coherence important for conductance and can reveal subtle
additional effects.
Such effects of interference between this coherence and the hyperspace
construction have also been presented.
It is interesting that most of the time, these give rise to exceedingly simple
behaviors.
\vspace{5pt}

Although we do not study the thermodynamic limit with any rigor, the finite $N$
results are
 $\it{locally}$ $\it{representative}$
because of the  self-similarity in the resistance v.s. length
behavior.
Our focus
 on the local behaviors around the special conducting
points $k_s$
is justified by the fact that these are the only points relevant to conduction
for $N >> 1$, so that our results on resistance patterns should be
representative
for big systems.
Some of these results could be tested experimentally in real chains
that can be manufactured rather easily with recent advances in
microfabrication techniques\cite{microfabrication}.

\section{figures captions}

FIG.1 Hyperspace construction of the standard Fibonacci chain (top)
and a $2/\tau$-shifted chain (bottom) ($2/\tau$ is the ``initial phase"
$\theta_0$
(in units where the width of the window is $1$) and it is equal to the
displacement of the physical line in the perpendicular direction).
The resulting chains, as well as the way that the defect is introduced,
are illustrated on the top left, through a step-modification of the
physical line.
\vspace{7pt}

FIG.2 (a) Typical local behavior of the Landauer resistance of the defected
chain (dashed line) compared to a pure chain (solid line), around a special
point.
The defected rises from zero  {\it lower} than the pure system.
(b) The same behavior but in a more extended region of $k$-space,
showing various crossovers, corresponding to chaotic behavior shown in
Fig. 8(c).
\vspace{7pt}

FIG.3 Comparison between resistances of a pure small chain ($N=34$) and all
possible defected chains of the same length, for a fixed $k$ close to a
special point ($k=k_s+\epsilon$). The $8$ possible defected chains are shown
on the bottom. We note a simple behavior with the defect-position ($x_P$)
and a resistance that is always lower than that of the pure chain
(horizontal line). This simple
(but discrete and asymmetric) behavior becomes smooth and symmetric for long
chains (see Fig. 4) where $x_P$ becomes quasi-continuous. $\;\epsilon=10^{-4}$
(in units $b=1$).
\vspace{7pt}

FIG.4 Corresponding results for long chains  $\;\;$(a) for $N=2000$
the simplest
possible pattern appears for $\phi_s={\pi \over N}$, and it is a smooth
and symmetric analog of Fig.3. This is always the pattern appearing for any $N$
(and $m=1$). It also turns out that all the defected chains are always less
resistive than the pure one (horizontal line). $\;\;\epsilon=10^{-9}$.
$\;\;$(b) Corresponding result for $m=5$. Again, a simple oscillatory pattern
appears (with the number of bumps equal to $m$), which is generally valid
for small $m$'s. These simple patterns are quasi-continuous analogs of
asymmetric and discrete patterns that are found for small chains,
as shown in (c) which corresponds to $N \simeq 200$.
\vspace{7pt}

FIG.5 Global view of the resistive behavior in $\phi_s$-line. See text for
details (end of Section 3). Shaded areas show regions of values of $\phi_s$
(and correspondingly of the potential strength parameter
 $\lambda$) where the resistive behavior is very
simple. The following three loops show the places where one recovers
simple resistive patterns and transitions in the resistive behavior.
\vspace{7pt}

FIG.6 Modulated resistive behaviors for $m$ between $N-4 P$ and $P$.
The defected system is, in the majority of cases, less resistive than
the pure one (horizontal line).
\vspace{7pt}

FIG.7 Recurrence of simple behaviors as $m$ crosses $P$ (for $N=2000$, that
yields  $P=472$ possible defect positions). Note a transition in the ordering
of resistances between defected and pure chains: solid curves correspond
to $m=P$; dashed to $m=P+1$; dot-dashed to $m=P+2$.
(Flat lines always give the values of the corresponding undefected chains).
\vspace{7pt}

FIG.8 Modulated and self-similar resistive patterns (close to special points)
for chains with length $N=3207$ (which yields $P=757$ possible defect
positions).
  (a) For $m=1603$, i.e. for $\phi_s$ in the first quadrant and just before
${\pi \over 2}$.
 (b) For $m=1605$, i.e. for  $\phi_s$ in the second quadrant and
 two units after ${\pi \over 2}$.
Both (a) and (b) are given for $\epsilon=10^{-4}$.
(c) As $\epsilon$ increases ($\epsilon=0.5$)  we recover chaotic behavior
(the values of the parameters are the same as in case (b)).

\end{document}